\documentstyle[letters]{mn}

\newif\ifAMStwofonts

\newcommand{\paper}{{Letter}}
\newcommand{\etal}{et al.}

\newcommand{\kms}{{\rm km\ s}^{-1}}

\ifoldfss
  \ifCUPmtlplainloaded \else
    \NewTextAlphabet{textbfit} {cmbxti10} {}
    \NewTextAlphabet{textbfss} {cmssbx10} {}
    \NewMathAlphabet{mathbfit} {cmbxti10} {} 
    \NewMathAlphabet{mathbfss} {cmssbx10} {} 
  \fi
  \ifAMStwofonts
    \ifCUPmtlplainloaded \else
      \NewSymbolFont{upmath} {eurm10}
      \NewSymbolFont{AMSa} {msam10}
      \NewMathSymbol{\upi}     {0}{upmath}{19}
      \NewMathSymbol{\umu}     {0}{upmath}{16}
      \NewMathSymbol{\upartial}{0}{upmath}{40}
      \NewMathSymbol{\leqslant}{3}{AMSa}{36}
      \NewMathSymbol{\geqslant}{3}{AMSa}{3E}

    \fi
  \fi
\fi 

\ifnfssone
  \newmathalphabet{\mathit}
  \addtoversion{normal}{\mathit}{cmr}{m}{it}
  \addtoversion{bold}{\mathit}{cmr}{bx}{it}
  \newmathalphabet{\mathbfit} 
  \addtoversion{normal}{\mathbfit}{cmr}{bx}{it}
  \addtoversion{bold}{\mathbfit}{cmr}{bx}{it}
  \newmathalphabet{\mathbfss} 
  \addtoversion{normal}{\mathbfss}{cmss}{bx}{n}
  \addtoversion{bold}{\mathbfss}{cmss}{bx}{n}
  \ifAMStwofonts
    \ifCUPmtlplainloaded \else
      \UseAMStwoboldmath
      \makeatletter
      \new@mathgroup\upmath@group
      \define@mathgroup\mv@normal\upmath@group{eur}{m}{n}
      \define@mathgroup\mv@bold\upmath@group{eur}{b}{n}
      \edef\UPM{\hexnumber\upmath@group}
      \new@mathgroup\amsa@group
      \define@mathgroup\mv@normal\amsa@group{msa}{m}{n}
      \define@mathgroup\mv@bold\amsa@group{msa}{m}{n}
      \edef\AMSa{\hexnumber\amsa@group}
      \makeatother
      \mathchardef\upi="0\UPM19
      \mathchardef\umu="0\UPM16
      \mathchardef\upartial="0\UPM40
      \mathchardef\leqslant="3\AMSa36
      \mathchardef\geqslant="3\AMSa3E
    \fi
  \fi
\fi 

\ifnfsstwo
  \DeclareMathAlphabet{\mathbfit}{OT1}{cmr}{bx}{it}
  \SetMathAlphabet\mathbfit{bold}{OT1}{cmr}{bx}{it}
  \DeclareMathAlphabet{\mathbfss}{OT1}{cmss}{bx}{n}
  \SetMathAlphabet\mathbfss{bold}{OT1}{cmss}{bx}{n}
  \ifAMStwofonts
    \ifCUPmtlplainloaded \else
      \DeclareSymbolFont{UPM}{U}{eur}{m}{n}
      \SetSymbolFont{UPM}{bold}{U}{eur}{b}{n}
      \DeclareSymbolFont{AMSa}{U}{msa}{m}{n}
      \DeclareMathSymbol{\upi}{0}{UPM}{"19}
      \DeclareMathSymbol{\umu}{0}{UPM}{"16}
      \DeclareMathSymbol{\upartial}{0}{UPM}{"40}
      \DeclareMathSymbol{\leqslant}{3}{AMSa}{"36}
      \DeclareMathSymbol{\geqslant}{3}{AMSa}{"3E}
    \fi
  \fi
\fi 

\ifCUPmtlplainloaded \else
  \ifAMStwofonts \else 
    \def\upi{\pi}
    \def\umu{\mu}
    \def\upartial{\partial}
  \fi
\fi

\title[The mass-temperature relation for clusters of galaxies]
  {The mass-temperature relation for clusters of galaxies}
\author[J. Hjorth, J. Oukbir and E. van Kampen]
  {Jens Hjorth$^{1,2}$\thanks{\mbox{E-mail: jens@nordita.dk (JH);}
  \mbox{jamila@dsri.dk (JO); eelco@tac.dk (EvK)}},
  Jamila Oukbir$^{3\textstyle\star}$ 
  and Eelco van Kampen$^{4\textstyle\star}$\\
  $^1$NORDITA, Blegdamsvej 17, DK--2100 Copenhagen \O, Denmark\\
  $^2$Centre for Advanced Study, Drammensveien 78, N--0271 Oslo, Norway\\
  $^3$Danish Space Research Institute, Juliane Maries Vej 30, 
      DK--2100 Copenhagen \O, Denmark\\
  $^4$Theoretical Astrophysics Center, Juliane Maries Vej 30, 
      DK--2100 Copenhagen \O, Denmark
  }
\date{\phantom{Accepted 1998 February \ \ . Received 1998 February \ \ ; 
in original form 1997 October 13}}
\pagerange{\pageref{firstpage}--\pageref{lastpage}}
\pubyear{1998}

\def\LaTeX{L\kern-.36em\raise.3ex\hbox{a}\kern-.15em
    T\kern-.1667em\lower.7ex\hbox{E}\kern-.125emX}

\begin{document}

\label{firstpage}

\maketitle

\begin{abstract}
A tight mass-temperature relation, $M(r)/r\propto T_X$, is expected in most 
cosmological models if clusters of galaxies are homologous and the intracluster 
gas is in global equilibrium with the dark matter. We here calibrate this 
relation using 8 clusters with well-defined global temperatures measured 
with {\it ASCA} and masses inferred from weak and strong gravitational lensing. 
The surface lensing masses are deprojected in accordance with N-body 
simulations and analytic results. The data are well-fit by the 
mass-temperature relation and are consistent with the empirical normalisation 
found by Evrard \etal\ (1996) using gasdynamic simulations. Thus, there is 
no discrepancy between lensing and X-ray derived masses using this approach.
The dispersion around the relation is 27 per cent, entirely dominated by 
observational errors. The next generation of X-ray telescopes combined with 
wide-field {\it HST} imaging could provide a sensitive test of the 
normalisation and intrinsic scatter of the relation resulting in a powerful 
and expedient way of measuring masses of clusters of galaxies. In addition, 
as $M(r)/r$ (as derived from lensing) is dependent on the cosmological model 
at high redshift, the relation represents a new tool for determination of 
cosmological parameters, notably the cosmological constant $\Lambda$.

\end{abstract}


\section{Introduction}

Clusters of galaxies are the largest gravitationally bound structures in 
the Universe and are as such excellent probes of cosmic structure formation 
and evolution. The ensemble properties of clusters expected in various 
cosmological scenarios can be used to derive constraints on the power spectrum 
of the initial density perturbations and on cosmological parameters such as 
$\Omega_0$ and $\Lambda$ (e.g., Eke, Cole \& Frenk 1996; 
Bahcall, Fan \& Cen 1997; Oukbir \& Blanchard 1997; Bartelmann \etal\ 1998;
de Theije, van Kampen \& Slijk\-huis 1998). On the scales of individual 
clusters the inferred baryon mass fraction can be used to constrain $\Omega_0$ 
(White \etal\ 1993; Evrard 1997). In such studies, an important quantity is 
the total cluster mass or any observed quantity which is tightly related to 
the mass. 

A promising mass estimator is the mean emission-weighted temperature, $T_X$, 
of the hot intracluster medium (ICM) in clusters of galaxies. Based on 
numerical simulations, it has been shown that $T_X$ is a better indicator of 
the total mass of a cluster than any other optical or X--ray property 
(Evrard 1990). Recently,
Evrard, Metzler \& Navarro (1996, hereafter EMN) and Eke, Navarro \& Frenk 
(1997) showed that there is a tight relation between the mass of a cluster and 
its global X-ray temperature in cosmological gasdynamic simulations, 
irrespective of the state of the cluster (e.g., not restricted to clusters 
with a `regular' appearance or  `isothermal' clusters) and the assumed 
cosmological model. In the simulations, it was found that mass predictions 
using this method (which only involve temperatures) are twice as precise 
as those derived using the $\beta$-model (which require the surface 
brightness distribution in addition, i.e., more photons and higher spatial 
resolution). However, the normalisation of the relation hinges on numerical 
simulations which may not comprise sufficient detail 
(EMN; Anninos \& Norman 1996). Therefore, it is essential to calibrate this 
relation from an observational point of view, by using independent mass 
estimators.

The purpose of this \paper\ is to provide a first observational calibration 
of the $M$--$T_X$ relation using the relatively `clean' way of determining 
independent cluster masses by gravitational lensing.
This technique essentially probes the projected mass 
along the line of sight. It is also pointed out that the relation holds the 
promise of providing a test of the geometry of the Universe which is 
particularly sensitive to $\Lambda$. Throughout this \paper, however, we 
assume a standard homogeneous Einstein--de Sitter Universe with $H_0=100 h$ 
km s$^{-1}$ Mpc$^{-1}$, $h=0.5$, $\Omega_0=1$ and $\Lambda = 0$.

\section{The mass-temperature relation}

Navarro, Frenk \& White (1997, hereafter NFW) found in their numerical 
simulations that the dark-matter distribution in present-day clusters have
self-similar density profiles when the radial coordinate is scaled to the 
radius containing an overdensity of $\delta=200$ relative to the critical 
density. More precisely, defining the overdensity as
\begin{equation}
\delta (r_\delta,z)\equiv {3M_\delta(r_\delta)\over 4\pi \rho_c(z) r_\delta^3},
\end{equation}
where $\rho_c(z)=\rho_{c0}(1+z)^3$ and $\rho_{c0}=3 H_0^2/(8 \pi G)$, NFW 
found that clusters are well-described by the density profile
$\rho (x) \propto x^{-1} (1+cx)^{-2}$, where $x=r/r_{200}$, in any cosmology. 
The variation in $c\approx 5$--10 with mass, cosmological parameters, and
redshift is small (Cole \& Lacey 1996; NFW; Bartelmann \etal\ 1998;
Eke, Navarro \& Frenk 1997) and so clusters form a homologous family to
a good approximation when scaled to a given overdensity. Optical
(Carlberg \etal\ 1997) and lensing observations (Fischer \& Tyson 1997) 
seem to support this conclusion.

For a cluster in quasi-equilibrium (Natarajan, Hjorth \& van Kampen 1997)
the virial theorem for the dark matter states that 
$ M(r) \propto r \left< v^2\right>_r$. Self-similarity
implies that the constant of proportionality 
depends on the adopted overdensity only. Finally, the global X-ray temperature 
is assumed to be proportional to the global mean velocity dispersion of 
the dark matter (at any time), i.e., $ T_X \propto \left< v^2\right>_r$. 
For example, this would be the case in the absence of transient effects and 
non-gravitational heating or cooling effects. In the case of equipartition one 
would have a universal $\beta\equiv \mu m_p \left< v^2\right>_r/(kT)=1$. 
Combining these assumptions (quasi-equilibrium, self-similarity, proportionality
between dark-matter velocity dispersion and X-ray temperature) leads to a 
simple scaling relation between the characteristic mass and radius at a given 
overdensity, $\delta$, and the global emission-weighted temperature of the hot 
X-ray gas,
\begin{equation}
M_\delta(r_\delta) = k_{\delta} r_\delta T_X,
\end{equation}
where $k_{\delta}$ is a constant depending on $\delta$. This equation 
expresses the structural invariance of clusters under mass and redshift 
transformations and {\em does not} rely on any particular dark-matter density 
profile or the assumption of hydrostatic equilibrium. 

Combined with the definition of the overdensity (eq.~1) this expression 
leads to the mass-temperature relation,
\begin{equation}
M_\delta =k_{\delta}^{3/2} \left ({3 \over 4\pi \delta \rho_{c0}}
\right )^{1/2} \left ({T_X\over 1+z}\right)^{3/2},
\end{equation}
or, equivalently, the size-temperature relation,
\begin{equation}
r_\delta (1+z) =k_{\delta}^{1/2} \left ({3 \over 4\pi \delta \rho_{c0}}
\right )^{1/2} \left ({T_X\over 1+z}\right)^{1/2},
\end{equation}
where $r_\delta (1+z)$ is the co-moving angular radius of the cluster.

Rather than trying to compute the prefactor $k_\delta$ from first principles, 
EMN used numerical simulations to calibrate these relations. They found the 
radius $r_{500}$ to be a conservative estimate of the boundary between the 
virialised region of the clusters and their outer envelopes. At $z=0.04$ 
using $\delta=500$ they found a universal prefactor independent of $\Omega_0$,
\begin{equation}
M_{500} = 2.22 \times 10^{15} 
  \left ( {T_X \over 10\ {\rm keV}}\right )^{3/2} {\rm M}_\odot.
\end{equation}
In the simulations the scatter around this relation was found to be only 15 
per cent compared to 30 per cent when using the $\beta$-model to estimate the 
mass.

Mohr \& Evrard (1997) have recently shown that observations of nearby clusters 
lead to an intrinsic scatter of 10--15 per cent in the relation between 
cluster isophotal size and mean emission-weighted temperature $T_X$ 
(similar to eq.~4) regardless of the state of the cluster 
(merging, cooling flow) thus giving added support to the existence of a tight 
mass-temperature relation.

\section{Lensing masses}

In order to test and calibrate the mass-temperature relation observationally 
we shall use independent masses determined from gravitational lensing. 
Since lensing masses are given in the literature as a function of physical 
radius rather than overdensity, we shall use eq.~(2) to express the 
temperature as a function of $M(r)/r$ instead of $M_{\delta}$. Thus the 
relation we shall test observationally is
\begin{equation}
\left ( {M(r) \over 10^{15}\ {\rm M_\odot}} \right )
\left  ({ 1\ {\rm Mpc} \over r} \right )
 = k_{\delta} 
\left  ({T_X \over 10\ {\rm keV}} \right ) .
\end{equation}
For an isothermal sphere, $M(r)\propto r$, $k_\delta$ would be a constant
independent of radius or overdensity. However, given the fact that clusters 
are described by more complicated density and temperature profiles 
(NFW; EMN) $k_\delta$ varies slightly with $\delta$ in the range considered. 
While an overdensity of $\sim 500$ was recommended ($r_{500}\sim$ 1--2 Mpc), 
EMN provided normalisations for $\delta=100,250,500,1000,2500$. 
Converting these into equivalent values for $k_\delta$ we find 
0.76,0.91,1.01,1.09,1.14. These slowly varying numbers are used to compute 
$k_\delta$ as a function of $\delta$ by spline interpolation. 

\subsection{Deprojection}

\begin{figure}
  \includegraphics{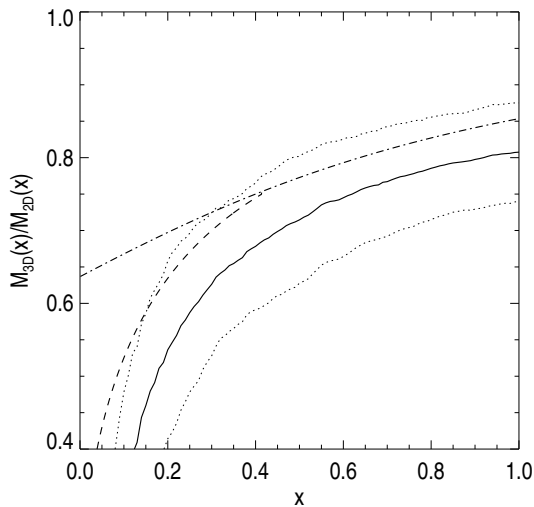}
  \vspace{7.2cm}
  \caption{This figure shows $M_{\rm 3D}(x)/M_{\rm 2D}(x)$ as a function of 
  $x=r/r_{200}$ for the Hernquist (1990) profile (dashed curve) and the model 
  of BBS (dashed-dotted curve). The solid curve is the corresponding mean
  deprojection factor for simulated massive clusters in a CDM $\Omega=1$ 
  Universe (see text for details) and the dotted curves indicate 
  the 1$\sigma$ confidence interval. 
  The corresponding curve for an open  CDM $\Omega_0=0.2 $ Universe (not
  plotted here) lies 10 per cent higher in good agreement with the Hernquist 
  model.}
  \label{fig1}
\end{figure}

Lensing provides the 2D projected (surface) mass, $M_{\rm 2D}(R)$ (where $R$ 
indicates a projected radius), of the cluster. In general the 2D mass at a 
given radius is larger than the 3D mass, $M_{\rm 3D}(r)$, evaluated 
at the same radius ($r=R$). 
The best way of obtaining a deprojection relation for $M_{\rm 3D}$ (which
is the quantity entering eq.~6) is to study numerical simulations of 
galaxy clusters, preferably a fair sample of these. We have used the catalogue 
of simulated standard CDM clusters ($\Omega=1$) of van Kampen \& Katgert (1997)
to find such a relation (van Kampen, in preparation). As the clusters we 
shall study 
in this \paper\ are biased towards massive clusters, we selected only clusters 
with a total mass within the Abell radius (3 Mpc) of at least 
$10^{15}$ M$_\odot$. The deprojection relation $M_{\rm 3D}(x)/M_{\rm 2D}(x)$ 
for these 41 clusters is plotted as a function of dimensionless radius 
$x=r/r_{200}$ in Fig.~1, along with its scatter, which is fairly substantial 
at small radii as substructure along the line of sight becomes important for 
the projected mass. In an open $\Omega_0=0.2$ CDM Universe the corresponding 
curve (not plotted) is 10 per cent higher due to the smaller influence of 
substructure.

In Figure~1 we also plot the deprojection relation of the Hernquist (1990) 
and the Brainerd, Blandford \& Smail (1996, hereafter BBS) models which 
both have analytic deprojection properties (the BBS model is the limit of 
$\eta\to\infty$ of the Hjorth \& Kneib (1998) model). As the total mass of the 
NFW model is infinite it is not useful for this purpose. However, we have made 
use of the 
fact that the half-mass radius in the Hernquist model is roughly equivalent to 
$r_{200}$ in the NFW model for $c\approx 5$ (Cole \& Lacey 1996). In the outer 
parts the Hernquist and BBS models coincide, but in the centre there is a 
marked difference between the two curves because of their differing divergence 
properties. The Hernquist model which diverges as $\rho\sim r^{-1}$ in the 
centre is similar to the NFW profile and $M_{\rm 3D}(x)/M_{\rm 2D}(x)\to 0$ 
for $x\to 0$ while the BBS model, which has a stronger central cusp 
$\rho\sim r^{-2}$, tends to the value for the singular isothermal sphere 
$2/\pi\approx 0.64$. This shows that deprojection of 2D masses at small radii 
depends sensitively on the exact slope of the inner cusp of dark-matter density 
profiles (Fukushige \& Makino 1997; Moore \etal\ 1998; Kravtsov \etal\ 1998). 
We finally note that the Hernquist model is in excellent agreement with the 
numerical results of the open model.

In this \paper\ we shall use the relation as a function of proper radius $R$
to deproject the lensing masses. For this purpose we 
introduce a convenient fitting function,
\begin{equation}
{M_{\rm 3D}\over M_{\rm 2D}} (R) = 0.56 \tan^{-1}\left ( {R\over 
0.28 {\rm Mpc}} \right ),
\end{equation}
where the coefficients have been determined from a non-linear least-squares
fit up to $R=2$ Mpc. We note that if such a deprojection correction is not 
applied, lensing (2D) masses will be higher than X-ray (3D) masses by a factor 
of 1.5 on average. 

\section{Data}

We have compiled a list of clusters with well-determined X-ray temperatures 
and masses determined independently using gravitational lensing. The data 
are shown in Table~1.

The X-ray data used here are from a recent compilation of temperatures 
of intermediate and high redshift clusters observed by the 
{\it Advanced Satellite for Cosmology and Astrophysics (ASCA)} 
(Mushotzky \& Scharf 1997). The temperatures were measured in a uniform way 
out to a radius of 3--6 arcmin depending on the redshift of the cluster. 

The lensing masses are from various studies of individual clusters, mostly 
using the `weak lensing' method pioneered by Kaiser \& Squires (1993), but 
also from variations in number counts of background galaxies 
(Broadhurst, Taylor \& Peacock 1995; van Kampen 1998). We included only
clusters with masses determined out to radii larger than 400 kpc to minimise 
deprojection and substructure effects from the central regions of the clusters. 
One cluster mass (MS 1358+62) was determined from wide-field {\it HST}\/ 
imaging. The normalisation of the mass of A2163 was adjusted in comparison 
with X-ray masses (derived from the $\beta$ model), i.e., this mass is not 
completely independent of the temperature (Squires \etal\ 1997).

\begin{table*}
\begin{minipage}{155mm}
 \caption{Observational data on clusters with {\it ASCA} temperatures and
 lensing masses. The temperatures are from Mushotzky \& Scharf (1997).
 The lensing masses are generally taken from the most recent publication of 
 a given cluster. For the data of Smail \etal\ (1995) we have assumed an 
 uncertainty of 40 \% in the masses and adopted the no-evolution model 
 for the redshift distribution of faint background galaxies. For the data of
 Squires \etal\ (1996ab,1997) we have estimated the masses inside 210\arcsec. 
 All masses and radii are computed assuming $h=0.5$,
 $\Omega_0=1$ and $\Lambda = 0$.}
 \label{table1}
 \begin{tabular}{@{}lllllllll}
Cluster &  $z$ & $T_X$                  &  $R$  &  $M_{2D}(R)$ & 
$M_{3D}(R)$          &  $\delta$  & $k_\delta$ & Reference               \\

        &      & (keV)                  & (Mpc) &  $(10^{14}$ M$_\odot)$& 
$(10^{14}$ M$_\odot)$&            &            &                         \\

Abell 2218     & 0.17 & $7.48^{+0.53}_{-0.41}$ & 0.80  &   9.4$\pm$1.7  & 
6.5$\pm$1.2    & 2714 & 1.14                   & Squires \etal\ (1996a)   \\
Abell 1689     & 0.18 & $9.02^{+0.40}_{-0.30}$ & 0.48  &  10.0$\pm$1.8  &
5.8$\pm$1.1    &11057 & 1.15                   & Taylor \etal\ (1998)     \\
Abell 2163     & 0.20 & $12.7^{+2.0}_{-2.0}$   & 0.90  &  13.0$\pm$10   &
9.2$\pm$7      & 2524 & 1.14                   & Squires \etal\ (1997)    \\
Abell 2390     & 0.23 & $8.90^{+0.97}_{-0.77}$ & 0.95  &  10$\pm$4      &
7.2$\pm$2.9    & 1551 & 1.14                   & Squires \etal\ (1996b)   \\
MS 1455.0+2232 & 0.26 & $5.45^{+0.29}_{-0.28}$ & 0.45  &   3.6$\pm$1.4  &
2.0$\pm$0.8    & 3859 & 1.14                   & Smail \etal\ (1995)      \\
MS 1358.4+6245 & 0.33 & $6.50^{+0.68}_{-0.64}$ & 1.00  &   4.4$\pm$0.6  &
3.2$\pm$0.4    &  468 & 1.00                   & Hoekstra \etal\ (1998)   \\
RX J1347$-$1145& 0.45 & $11.37^{+1.10}_{-0.92}$& 2.00  &  34$\pm$8      &
27$\pm$6       &  385 & 0.98                   & Fischer \& Tyson (1997)  \\
MS 0015.9+1609 & 0.54 & $8.0^{+1.0}_{-1.0}$    & 0.60  &   8.5$\pm$3.4  &
5.4$\pm$2.2    & 2355 & 1.14                   & Smail \etal\ (1995)      \\

 \end{tabular}
\end{minipage}

\end{table*}

We show the results for the eight clusters in Fig.~2 in which we plot 
$M_{\rm 3D}(R)/R$ derived from lensing studies as a function of 
$k_\delta T_X$. The relation predicted by eq.~(6) gives an excellent fit to 
the data. The best-fit line has a normalisation which is 88 per cent of that 
predicted by EMN and 
the dispersion (rms) about the relation is 27 per cent in mass, somewhat 
smaller than that expected from the quoted observational errors alone. 

\begin{figure}
  \includegraphics{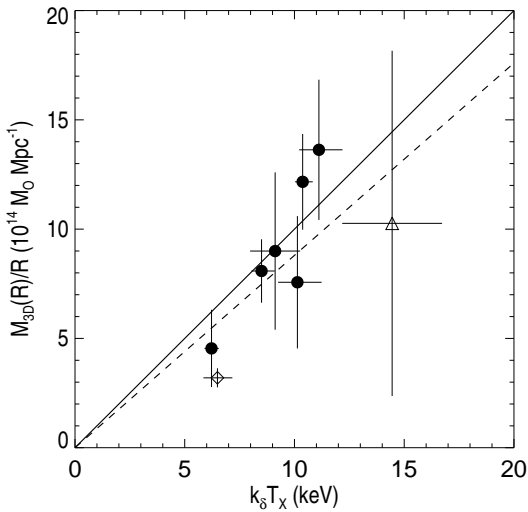}
  \vspace{7.2cm}
  \caption{This figure shows the $M_{3D}(R)/R$--$T_X$ relation for lensing
clusters of galaxies. Filled circles are ground-based data, the open
diamond is the {\it HST} data point of MS 1358+62, the open triangle
is A2163. The error bars do not include deprojection errors. The solid line 
is the relation predicted by eq.~(6) as normalised by EMN and has not been 
fitted to the data. The dashed line minimises the mean relative residual and 
has a normalisation 12 per cent lower.}
  \label{fig2}
\end{figure}

\section{Discussion}

The observed scatter around the mass-temperature relation (eq.~6) is dominated 
by observational errors and is consistent with having no intrinsic scatter. 
The data thus support the existence of an $M$--$T_X$ relation as a fairly 
accurate independent estimator of cluster masses.
It is, however, important to point out that the data set presented here may
be affected by systematic errors in both the masses and the temperatures.
Adopting different data sets could lead to significant changes in e.g.\
the normalisation of the mass-temperature relation.

X-ray temperatures of hot clusters are usually uncertain due to the few 
photons detected above 8 keV with {\it ASCA} and quoted errors normally do 
not incorporate possible systematic errors. Thus, 
many compilations (e.g., Sadat, Blanchard \& Oukbir 1998) may be affected by
the fact that temperatures often differ from author to author 
(Mushotzky \& Scharf 1997; Allen 1998; Yamashita 1997) due to
differences in data analysis and use of data from 
other satellites ({\it ROSAT, Ginga}). For example, Allen (1998) has shown 
that cooling flows may bias global temperatures as derived by 
Mushotzky \& Scharf (1997) downward by about 30 \% on average. In this \paper\
we have used the Mushotzky \& Scharf (1997) data because of the uniformity and 
simplicity of the analysis and the possibility of extending it to fainter
and higher redshift clusters.

Ground-based lensing data are affected by fairly large seeing 
corrections. {\it HST} results are therefore preferable but usually only 
available inside a small radius. The normalisation of weak-lensing masses 
(due to the 
mass-sheet degeneracy) can be carried in various ways, using e.g.\ multiple
arcs, magnification bias, fits of analytic models or comparison with 
X-ray profiles. Sadat \etal\ (1998) found that a normalisation of $k_\delta$
about 64 per cent of the EMN value (no deprojection was applied)  was 
consistent with the temperatures and 
masses of 5 {\it HST} clusters studied out to 400 kpc by Smail \etal\ (1997). 
Such a small normalisation would imply that the data presented here have 
systematically overestimated masses or underestimated temperatures. 
The high-quality {\it HST} data point for MS 1358+62 may indicate that
ground-based masses are indeed overestimated. However, the mass of
MS 1358+62 as derived from weak lensing could also be underestimated e.g.\
due to deprojection errors arising because of the high ellipticity of the
cluster. Such an
underestimate of the 3D mass is supported by the disagreement between the 
velocity dispersion derived from weak lensing ($780\pm50\ \kms$) and that
found from direct spectroscopic measurements as well as from strong lensing
($\sim 1000\ \kms$) (Hoekstra \etal\ 1998). {\it HST} clusters may also have 
underestimated masses due to the fact that a fit of an isothermal sphere to the
mean tangential shear inside a small radius (Smail \etal\ 1997) in general 
biases masses low (van Kampen \& Hjorth, in preparation).

It is a long standing discussion whether masses determined from lensing agree 
with or exceed X-ray masses determined using the $\beta$ model 
(see e.g.\ Smail \etal\ 1997; Allen 1998). If we take the results
presented in Fig.~2 at face value the good agreement between the EMN 
normalisation and the observational calibration indicates that there is no 
such discrepancy when using the mass-temperature relation. If anything the 
X-ray masses computed using the EMN normalisation are slightly higher 
(by $\sim$10--20 per cent) 
(cf.~Fig.~2 and MS 1358+62) than lensing masses. Such an effect would be
consistent with the predictions of simulations incorporating the effects of 
galactic winds (Metzler \& Evrard 1998) which contribute to heating the ICM. 

Besides its use as a straightforward mass-estimator for any cluster with 
a well-determined global temperature the mass-temperature relation holds the 
promise of becoming an important cosmological tool, bearing a resemblance with 
the Fundamental Plane or Tully--Fisher scaling relation for elliptical or 
spiral galaxies, respectively, in that it relies on simple scaling relations 
with 10--20 per cent scatter, but presumably involves much smaller evolutionary 
corrections.
A direct cosmological application of the $M$--$T_X$ relation would be to 
examine the inferred deviations from it as a function of redshift. A possible 
trend with redshift could be indicative of (i) evolutionary effects (ii) 
the assumed redshift distribution of the faint background galaxies, $N(z)$, 
or (iii) the parameters entering the assumed cosmological model.

Typical evolutionary effects could be non-gravitational 
heating or cooling of the ICM such as effects of feedback mechanisms like 
galactic winds which introduce systematic structural changes of the ICM
(Metzler \& Evrard 1998) or cooling flows (Allen 1998). Possible `outliers' 
from the relation could be due to e.g.\ merging clusters with a very unsettled 
temperature distribution (Schindler 1996) or highly elongated clusters which 
give large deprojection uncertainties depending on the viewing angle.

While the inferred lensing masses of low and intermediate redshift 
clusters are fairly insensitive to the assumed median 
redshift of the background galaxies, high-redshift clusters are very sensitive 
to the assumed median redshift (Smail \etal\ 1995; Luppino \& Kaiser 1997) and 
so deviations from the expected relation at high redshift could be used to 
constrain $N(z)$.

Finally, the world model enters through the derived masses and sizes via the 
expression for the angular diameter distance. In the simplest form, 
$T_X\propto D_S/D_{LS}$, where $T_X$ is a directly measurable intrinsic 
quantity and $D_S/D_{LS}$ is the ratio between the source and lens--source 
angular diameter distances. Thus, the method can be used as a test for the 
geometry of the Universe, which is less sensitive to inhomogeneities along the 
line of sight than small standard rods/candles (e.g., SN Ia) 
(Hadrovi\'c \& Binney 1997). Individual massive high-redshift clusters could 
therefore be fairly unbiased discriminators between different cosmological 
models, particularly sensitive to the cosmological constant $\Lambda$. At 
$z=1$ the difference between a $(\Omega_0,\Omega_\Lambda)=(1,0)$ and a 
$(\Omega_0,\Omega_\Lambda)=(0.2,0.8)$ Universe is 25 per cent in $D_S/D_{LS}$. 
Moreover, if the measurement of the global X-ray temperature is supplemented
with spatially resolved X-ray imagery additional constraints on $\Omega_0$ can 
be derived (Sasaki 1996; Pen 1997).

\section{Conclusion}

Based on numerical simulations (EMN; Eke \etal\ 1997) and observations of 
nearby clusters (Mohr \& Evrard 1997) the existence of a tight mass-temperature 
relation has been suggested. The results presented here provide support for 
this assertion and indicate that the mass-temperature relation (eq.~6) can be 
used to determine cluster masses with a precision of 27 per cent (Fig.~2).
There seems to be no significant discrepancy between deprojected lensing 
masses and masses derived from X-ray temperatures, using the normalisation 
found in numerical simulations (EMN). 

The origin of this tight relation is believed to be the fairly simple physics 
entering the relation (cf.~Sec.~2), namely virialisation of gravitationally 
bound structures with self-similar dark-matter density distributions that are 
in global quasi-equilibrium with the hot ICM, independent of the chosen world 
model, power spectrum, or exact formation redshift of the cluster.

We have cautioned that the observational data discussed in this Letter are 
quite uncertain and possibly affected by systematic errors. The results should 
therefore only be taken as an indication of a tight mass-temperature relation. 
However, the future observational situation is promising. A sample of clusters 
with very precise lensing masses (e.g., from wide-field {\it HST} imaging with 
the ACS) to about 10 per cent or better (e.g., Natarajan \etal\ 1998; 
Hoekstra \etal\ 1998) and equally accurate temperatures 
(e.g., with {\it AXAF}, {\it Spectrum--XG}, or {\it XMM}) would allow us to 
study the intrinsic scatter of the relation and determine a precise 
normalisation. This could provide a direct and reliable mass estimator for 
distant clusters with important cosmological implications.

\section*{ACKNOWLEDGMENTS}
We thank 
Monique Arnaud,
Henk Hoekstra,
Jean-Paul Kneib, 
and Ian Smail (the referee),
for useful comments and discussions.
JH acknowledges the hospitality of DSRI where part of this
work was carried out. 
This work was supported in part by Danmarks Grundforskningsfond 
through its funding of the Theoretical Astrophysics Center.



\label{lastpage}

\end{document}